\documentclass[12pt]{iopart}
\usepackage{iopams}

\usepackage[usenames,dvipsnames]{xcolor}
\usepackage{todonotes}
\usepackage[normalem]{ulem}

\newcommand{\hint}[1]{\texttt{\small [#1]}}

\makeatletter
\DeclareRobustCommand{\text}{%
  \ifmmode\expandafter\text@\else\expandafter\mbox\fi}
\let\nfss@text\text
\def\text@#1{{\mathchoice
  {\textdef@\displaystyle\f@size{#1}}%
  {\textdef@\textstyle\f@size{#1}}%
  {\textdef@\textstyle\sf@size{#1}}%
  {\textdef@\textstyle \ssf@size{#1}}%
  \check@mathfonts
  }%
}
\def\textdef@#1#2#3{\hbox{{%
                    \everymath{#1}%
                    \let\f@size#2\selectfont
                    #3}}}
\makeatother

\usepackage{graphicx}

\usepackage[utf8x]{inputenc}
\usepackage[OT1]{fontenc}

\usepackage{cite}
\usepackage{lmodern}
\usepackage{microtype}
\usepackage{textcomp}

\usepackage{bm}

\usepackage{pgfplots}

\usepackage{sansmath}
 \newcommand*\circled[1]{\tikz[baseline=(char.base)]{
 \node[shape=circle,draw,inner sep=1pt] (char) {#1};}}

\begin{document}

\title[A high-flux BEC source for mobile atom interferometers]{A high-flux BEC source for mobile atom interferometers}

\author{Jan Rudolph$^{1}$, Waldemar Herr$^{1}$, Christoph Grzeschik$^{2}$, Tammo Sternke$^{3}$, Alexander Grote$^{4}$, Manuel Popp$^{1}$, Dennis Becker$^{1}$, Hauke Müntinga$^{3}$, Holger Ahlers$^{1}$, Achim Peters$^{2}$, Claus Lämmerzahl$^{3}$, Klaus Sengstock$^{4}$, Naceur Gaaloul$^1$, Wolfgang Ertmer$^1$ and Ernst M. Rasel$^1$}

\address{$^1$ Institut f\"ur Quantenoptik, Leibniz Universit\"at Hannover, Welfengarten 1, 30167 Hannover, Germany}
\address{$^2$ Institut f\"ur Physik, Humboldt-Universit\"at zu Berlin, Newtonstraße 15, 12489 Berlin, Germany}
\address{$^3$ ZARM, Universit\"at Bremen, Am Fallturm, 28359 Bremen, Germany}
\address{$^4$ Institut f\"ur Laser-Physik, Universit\"at Hamburg, Luruper Chaussee 149, 22761 Hamburg, Germany}

\ead{rudolph@iqo.uni-hannover.de}


\begin{abstract}
Quantum sensors based on coherent matter-waves are precise measurement devices whose ultimate accuracy is achieved with Bose-Einstein condensates (BEC) in extended free fall. This is ideally realized in microgravity environments such as drop towers, ballistic rockets and space platforms. However, the transition from lab-based BEC machines to robust and mobile sources with comparable performance is a challenging endeavor. Here we report on the realization of a miniaturized setup, generating a flux of $4 \times 10^5$ quantum degenerate $^{87}$Rb atoms every 1.6\,s. Ensembles of $1 \times 10^5$ atoms can be produced at a 1\,Hz rate. This is achieved by loading a cold atomic beam directly into a multi-layer atom chip that is designed for efficient transfer from laser-cooled to magnetically trapped clouds. The attained flux of degenerate atoms is on par with current lab-based BEC experiments while offering significantly higher repetition rates. Additionally, the flux is approaching those of current interferometers employing Raman-type velocity selection of laser-cooled atoms. The compact and robust design allows for mobile operation in a variety of demanding environments and paves the way for transportable high-precision quantum sensors.
\end{abstract}

%
%
%
%
%

\section{Introduction}\label{sec:intro}
One of the major quests in modern physics is to unify the fundamental interactions of nature. However, none of the existing theories is widely considered successful so far. Several models, e.g.\ loop quantum gravity and string theory, that describe gravity within the same formalism as the other interactions, make quantitative predictions that can be tested experimentally\cite{Overduin2009}. A crucial feature of these theories is the prediction of violations of general relativity (GR) postulates at different accuracy levels, allowing for experimental tests to discriminate them. Einstein's equivalence principle (EEP) with its three pillars Lorentz invariance, local position invariance and the weak equivalence principle (WEP), is a promising candidate for such tests. Gravitational wave (GW) detection is another example where precision measurements are expected to test the predictions of GR on a new level\cite{Sathyaprakash2009}. However, all such endeavors require a new generation of measurement devices that feature the accuracy to probe for such violations.

In recent years, an excellent degree of control over ensembles of ultra-cold atoms has been demonstrated. By exploiting the wave character of atomic ensembles that is dominant on such temperature scales, atoms have been used as highly sensitive probes for a variety of physical quantities such as the gravitational constant G\cite{Fixler2007,Lamporesi2008} and the fine-structure constant\cite{Bouchendira2011}, as well as the measurement of accelerations\cite{Peters1999,Debs2011,Louchet2011} and rotations\cite{Durfee2006,Stockton2007,Gauguet2009,Takase2008,Tackmann2012}. While the accuracy of atom interferometers has improved to compete with classical tests, they also serve as complementary tools, as they may explore new physics and the fundamental structure of matter in the quantum regime.

The precision of atom interferometers can scale with the square of the free-fall time, suggesting a substantial advantage for large fountains\cite{Dickerson2013, Hartwig2015} and microgravity operations\cite{Geiger2011,Rudolph2011,Muentinga2013}. Space platforms are expected to push the performance of these devices to their ultimate limit\cite{Hogan2011, Altschul2015, Aguilera2014}. Besides the long interrogation times possible on a spacecraft, they provide a quiet, seismic noise-free platform with large variations in altitude, velocity and gravitational potential. All of these ingredients are expected to make the above-mentioned tests of fundamental laws very accurate. The next decisive step towards unfolding the true potential of ultra-cold atoms is hence to realize compact and robust devices that can be used in mobile applications such as gravimetry, geodesy, geophysics, seismology, navigation and tests of fundamental physics beyond the lab environment. Proof-of-principle atom interferometry experiments in microgravity have been demonstrated in 0\,$g$ parabola flights\cite{Geiger2011} and in a drop tower facility\cite{Zoest2010, Muentinga2013}.

State-of-the-art quantum sensors typically operate with laser-cooled atomic ensembles at temperatures of few \textmu K and cycle rates of about 1\,Hz. However, extending the free fall time for high-precision atom interferometry requires much lower temperatures. In this respect, delta-kick cooled BECs are of large interest, as they allow to obtain both low expansion rates and low densities, evading the mean field effects. Usually, Bose-Einstein condensation dramatically increases the preparation times to several tens of seconds, results in a low flux of atoms and complicates the design of the experiment, compromising its portability.

The BEC machine described in this paper reconciles a high flux of quantum degenerate atoms, compactness and portability. This is achieved by loading a pre-cooled atomic beam from a two-dimensional magneto-optical trap (2D$^{+}$MOT\cite{Dieckmann1998}) directly into a multi-layer atom chip. This novel technique not only eliminates the need for a standard MOT, employing macroscopic coil arrangements from where the atoms are subsequently loaded into the atom chip, but demonstrates atom numbers competitive with large, conventional assemblies as well as velocity-filtered thermal sources. Previous experiments loading an atomic beam into a chip trap have not led to high numbers of captured or condensed atoms\cite{Roux2008}. In this paper, we demonstrate the synergy between a high flux source with tunable velocity profile and a compact chip design with characteristically low capture volume and velocity.

With its three layers, the atom chip features an abundance of possible trap configurations, from shallow confinement for optimal loading, to tight traps for fast evaporation, and an efficient transfer between them. The capability to optimize each experimental phase allows us to achieve both high numbers of captured atoms and rapid evaporative cooling in a compact setup with low power consumption. The assembly is built to withstand forces of up to 50\,$g$ and thereby is suitable for field operation as well as deployment on microgravity platforms such as drop towers. The concept can easily be adapted for use on ballistic rockets and satellites, making it the baseline for a space mission payload\cite{Schuldt2014}.

This paper is organized as follows: First, we introduce the experimental setup with a focus on the vacuum chambers and the atom chip. The experimental methods and performance is presented in Section \ref{sec:results}. The performance limiting factors of our current setup are assessed using a theoretical model of the evaporation process. In Section \ref{sec:comparison}, the results in comparison to other fast BEC experiments, both lab-based and mobile, as well as possible improvements are discussed. In the conclusion, we recall the main results and highlight application perspectives for the developments presented in this paper.

\section{Experimental Setup}\label{sec:setup}

\begin{figure}[t]
	\centering
		\includegraphics[width=0.43\textwidth]{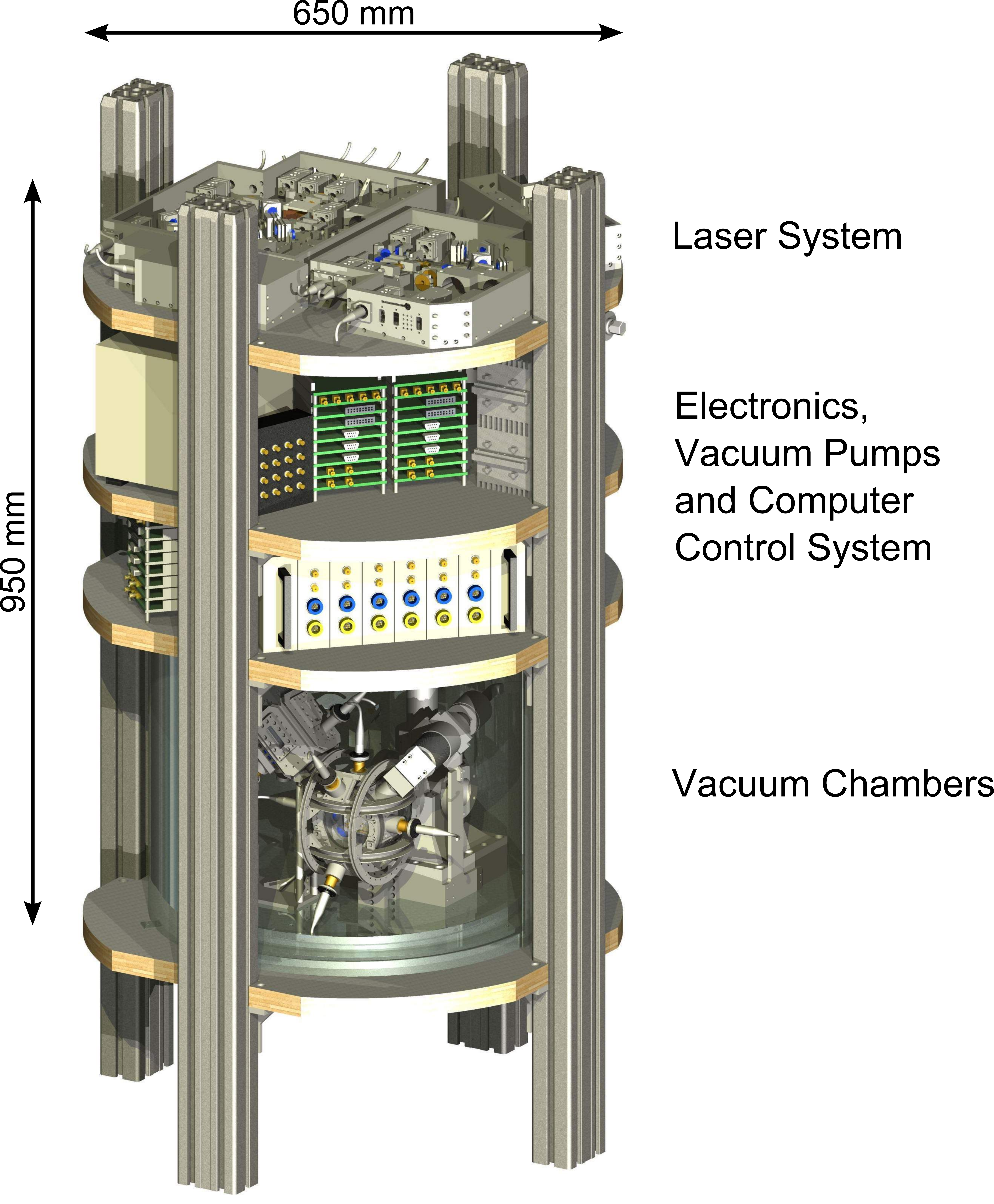}
	\caption{CAD model of the experimental setup. The assembly consists of four platforms, each with a diameter of approximately 65\,cm. From top to bottom, they are used for the laser system, vacuum pumps, computer control system and electronics, and the vacuum chambers.}
	\label{fig:capsule}
\end{figure}

In this section, we describe the apparatus with focus on the most relevant components of the atom chip based assembly illustrated in Figure \ref{fig:capsule}. The total weight of the setup is 245.8\,kg, 98.8\,kg of which is owed to the application specific structure (stringers and platforms). Hence, the payload mass amounts to 147\,kg including a two layer MuMetal shield around the vacuum chambers (42.6\,kg). The footprint of the assembly could be reduced further using a different support structure. The power consumption of the entire setup in operation is 363.9\,W. Hence, it can easily be run on commercially available accumulators for several hours at a time.

\subsection{Laser System}
Ridge waveguide laser diodes \hint{Eagleyard EYP-RWE-0780-02000} in a compact external cavity diode laser configuration\cite{Baillard2006,Gilowski2007} generate cooling and repumping light for the $^{87}$Rb MOT. A master laser is frequency stabilized to the $D_2$ line of the rubidium spectrum via saturated absorption spectroscopy, while two other lasers are subsequently stabilized via frequency offset locks. The cooling light is tuned a few linewidths to the red of the $F=2 \rightarrow F'=3$ transition, while the repumping light is resonant to $F=1 \rightarrow F'=2$.

The cooling and repumping light are mixed on two beam splitting cubes and amplified in two tapered amplifiers (TA) \hint{m2k TA-0780-1000}, one for 2D and one for 3D-MOT operation. Each TA generates up to 1\,W of output power in a cooling to repumping ratio of \textgreater 10:1. The light is delivered to the vacuum chambers via polarization maintaining single-mode optical fiber splitters \hint{OZ-Optics FOBS-14P-11111-5/125-PPPPP-770-45/45/9/1, Canadian Instruments 954P fixed ratio pmcoupler}. The total light power available for 2D$^{+}$MOT operation is 120\,mW, while the 3D-MOT is operated with a total power of 92\,mW. A small part of about 2\,mW of the cooling light is split off and used for optical state preparation via the $F=2 \rightarrow F'=2$ transition as well as for fluorescence and absorption imaging via the cooling transition.

\subsection{Vacuum Chambers}
\begin{figure}
	\centering
		\includegraphics[width=1.0\textwidth]{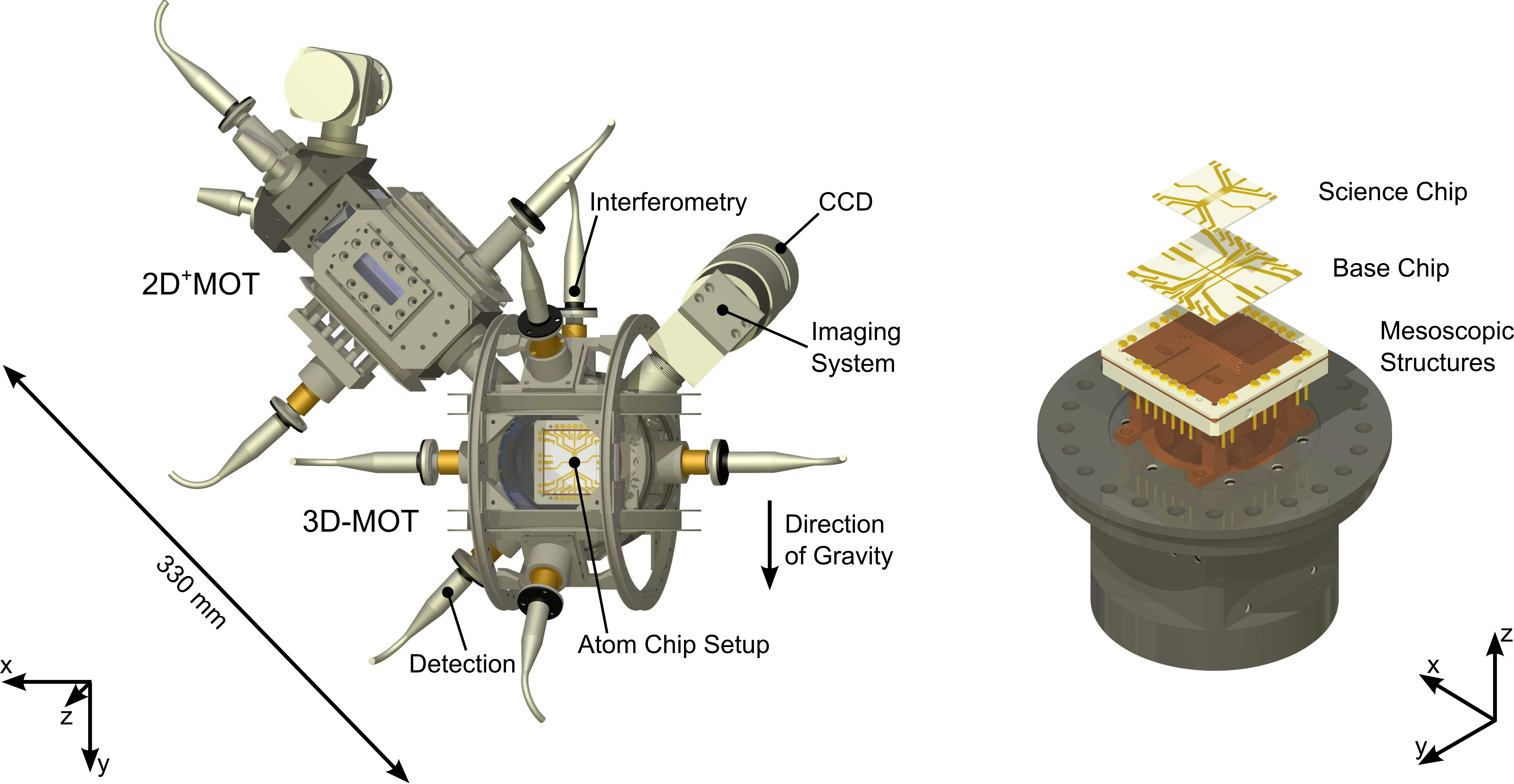}
	\caption{CAD models of the vacuum setup (left) and the atom chip setup (right). The vacuum setup consists of two chambers. In the 2D chamber a high rubidium background pressure is created to form a 2D$^{+}$MOT\cite{Dieckmann1998}, which generates a beam of pre-cooled atoms. This beam is injected into the 3D chamber, where an ultra-high vacuum is maintained. Here, the atoms are collected by a 3D-Chip-MOT and transferred to pure magnetic traps, formed by the atom chip setup and external bias coils. Atom interferometry as well as detection of the atoms are carried out in the 3D chamber. On the right hand side the three layers of the atom chip setup, consisting of the science chip, the base chip and the mesoscopic structures, are shown in an exploded view.}
	\label{fig:chambers}
\end{figure}

The vacuum setup consists of two chambers separated by a differential pumping stage (see Figure \ref{fig:chambers}), allowing for a pressure difference of up to three orders of magnitude. A high vacuum (HV) area (2D chamber) is used for the atomic source and is operated slightly below the room temperature vapor pressure of rubidium at $10^{-7}$\,mbar. It generates a pre-cooled beam of atoms towards an ultra-high vacuum (UHV) chamber. The UHV region (3D chamber) is used to capture the atoms, cool them to degeneracy and perform atom interferometry. Its pressure level is maintained at \textless $10^{-10}$\,mbar by a 25\,l/s ion pump \hint{IGP Meca 2000} and two passive vacuum pumps \hint{VG Scienta SBST110, SAES Getters CapaciTorr D200}. 

The chambers are machined from a non-magnetic Titanium alloy \hint{Ti-6AL-4V}. The differential pumping stage between the two is a threaded copper rod with an 1.5\,mm aperture in the center, which after 10\,mm expands conically with an aperture angle of 8\textdegree{} for another 30\,mm. The conical part is partially replaced by a graphite tube to improve differential pumping. On the HV side, the pumping stage has a 45\textdegree{} cutaway with a polished surface that grants upwards of 95\% reflectivity at a wavelength of 780\,nm.

\subsubsection{2D Chamber}
The 2D chamber has a cuboid shape with a head section that holds three CF-10 vacuum ports for dispensers, a CF-16 vacuum port to access a rubidium reservoir and a window \hint{N-BK7} for optical access along the main axis of the chamber. The inner (outer) dimensions of the 2D chamber are $6\,\text{cm} \times 2\,\text{cm} \times 2\,\text{cm}$ ($14\,\text{cm} \times 7.4\,\text{cm} \times 7.4\,\text{cm}$).

The lateral chamber windows \hint{N-BK7} are anti-reflection (AR) coated on the outside and attached to the chamber via Indium sealing. Each lateral axis has one linearly polarized input beam that is expanded to a diameter of 18\,mm, split into two parallel beams and transformed to circular polarization. After passing the chamber, rectangular $\lambda/4$ waveplates, which are AR coated on one and high-reflection (HR) coated on the other side, retro reflect the beams to create two cooling regions along the longitudinal axis. Additionally, a pushing and a retarding beam are counter-propagating along the longitudinal axis to create a 2D$^{+}$MOT configuration. The retarding beam is reflected off the differential pumping stage and hence has a 1.5\,mm cutaway in its center.

The magnetic fields for the 2D$^{+}$MOT are provided by four coils in racetrack configuration, generating a two-dimensional quadrupole field perpendicular to the atomic beam.

\subsubsection{3D Chamber}
The 3D chamber has a cylindrical shape with outer dimensions of $\text{\o}\,102\,\text{mm} \times 62\,\text{mm}$. It has eight viewports on its lateral surface, seven of which have Indium-sealed windows \hint{N-BK7} with AR coatings on both sides. The window sizes grant a free aperture of 20\,mm. The remaining port holds the differential pumping stage towards the 2D chamber. The axis perpendicular to the differential pumping stage is used for absorption imaging. The atoms are illuminated from the lower left viewport and detected via a CCD camera \hint{Hamamatsu C8484-15C} that sits behind a two-lens imaging system on the upper right. The horizontal axis is used for a pair of MOT beams while the vertical axis is used for atom interferometry.

The chamber features a large front window that grants optical access for two more MOT beams. Each of these beams enters the chamber in an angle of 45{\textdegree} to the plane of the window and is reflected by the atom chip\cite{Reichel1999}. A large aperture lens system \hint{Thorlabs MAP105050-B} sits in front of the window, collecting light emitted by the atoms onto a photodetector \hint{Hamamatsu S5107}.

Three pairs of Helmholtz coils are attached to the outside of the 3D chamber, one of which is wound around the chamber itself. The chip setup constitutes the back side of the assembly and provides access to the vacuum pumps.

\subsection{Atom Chip Setup}
The magnetic fields for trapping the atoms are created by current carrying wire structures in combination with magnetic bias fields\cite{Reichel1999}. Three layers of wire structures, each featuring a different characteristic wire size, are used in the atom chip setup.

The first layer holds the largest, mesoscopic structures which are constructed from Kapton isolated 0.9\,mm diameter copper wires. These are used in the generation of the quadrupole field for the 3D-MOT with a U-shaped layout\cite{Wildermuth2004} that comprises six windings of a single wire. Additionally, three individual copper wires form an H-shaped structure to generate a Ioffe-Pritchard (IP) type potential, that is used in the first magnetic trap.

The second layer (base chip) features intermediate sized gold wires of 0.5\,mm width, electroplated onto a 35\,mm $\times$ 35\,mm Aluminum nitride substrate. A 25\,mm $\times$ 25\,mm chip (science chip) forms the third and final layer with structures of 50\,\textmu m width. It is covered with a dielectric transfer coating \hint{OIB Jena} to reflect two of the four MOT beams, creating a mirror MOT configuration\cite{Reichel1999}. Its reflectivity at 780\,nm was measured to be 97.7\% at a 45° angle of incidence. The base and science chip feature a set of 4 and 5 parallel wires, respectively, in each case intersecting with one central orthogonal wire. They both offer an abundance of possible U-, Z- and H-shaped trap configurations including dimple traps\cite{Reinhard2009}.

Previous experiments have used larger, mesoscopic structures to increase the number of captured atoms\cite{Wildermuth2004}. However, the ability to create high trap frequencies with such structures is limited for two reasons. First, the range of feasible currents that can be applied to the structures is limited. Second, the high trap frequencies necessary for fast evaporation can only be attained in close vicinity to the structure. The exclusive use of such mesoscopic structures, therefore, has led to slow evaporation. Other experiments, that employ only small chip structures typically show fast evaporation performances but are highly limited by the initial number of atoms\cite{Zoest2010}. Our three layer chip is designed to bridge the gap between these two scenarios by using traps composed of different chip layers to span a wide range of trap configurations (see Figure \ref{fig:loading}).

The entire atom chip setup is located inside the UHV environment and was assembled using low-outgassing UHV suitable epoxy adhesives \hint{Epotek 353ND, H77}. The currents of up to 10\,A are provided by battery supplied current drivers \hint{High Finesse BCS 10A}, which are rated at a noise level below -108\,dB V$_\text{RMS}$ between 0 and 2500\,Hz.
\section{Methods and Results}\label{sec:results}
\begin{figure}[t]
	\centering
		\includegraphics[width=1.0\textwidth]{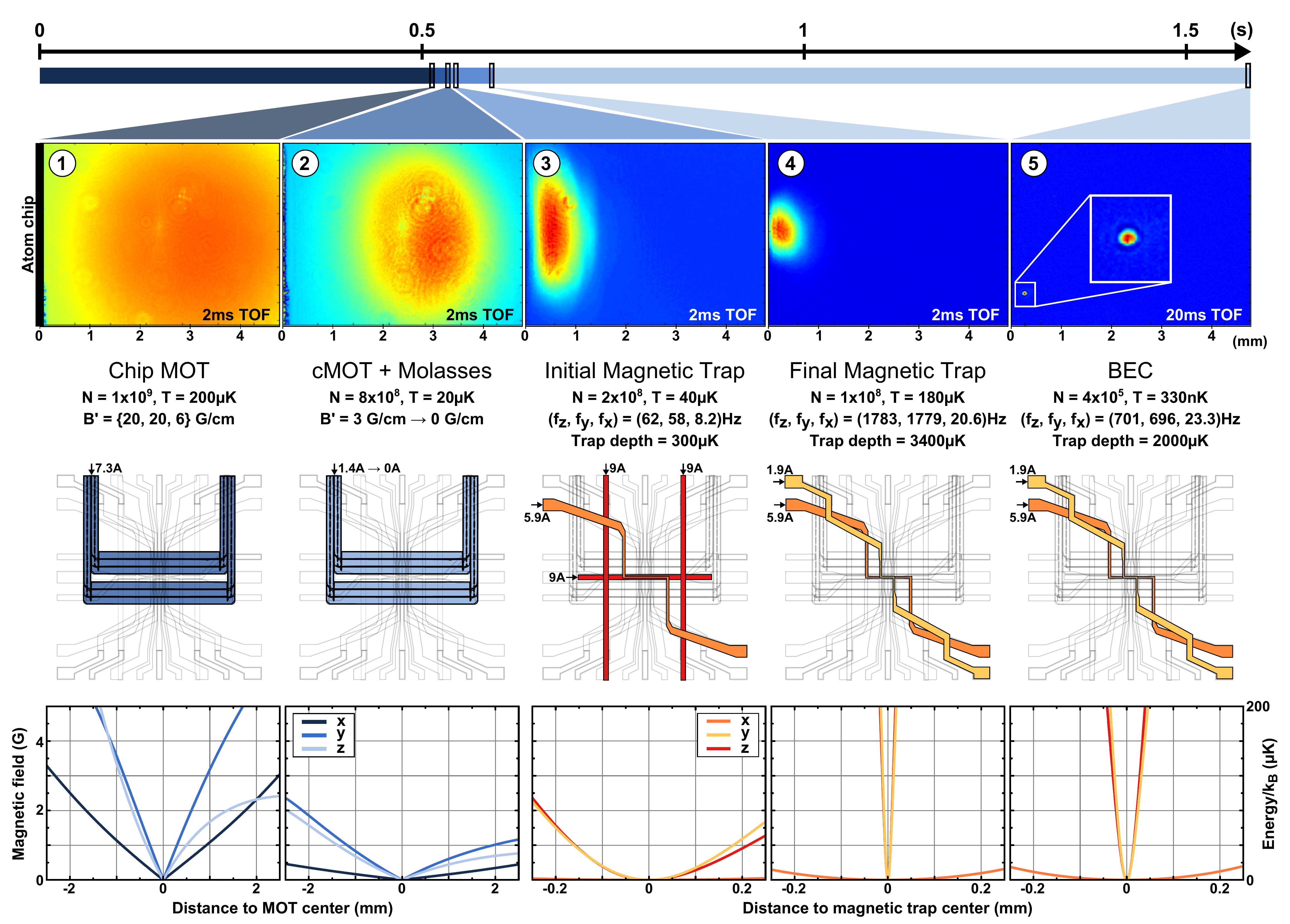}
	\caption{Source scheme to prepare 4 $\times$ 10$^5$ quantum degenerate atoms in 1.6\,s. Five absorption images of the atoms illustrate the steps involved (\scriptsize\circled{1} - \circled{5}\small). The chip structures used as well as the magnetic field calculated with a model of the wire structures are shown below the images. (The trap bottom has been substracted for the magnetic traps.) All chip configurations are used in conjunction with external bias fields.\;\scriptsize\circled{1}\small\; After 500\,ms 1 $\times$ 10$^9$ atoms are loaded into a MOT generated by the mesoscopic U structure.\;\scriptsize\circled{2}\small\; The atoms are compressed and molasses cooled to 20\,$\mu$K.\;\scriptsize\circled{3}\small\; 2 $\times$ 10$^8$ atoms can be captured in the initial magnetic trap, formed by the mesoscopic H and a base chip Z structure.\;\scriptsize\circled{4}\small\; To improve the evaporation efficiency, the trap is compressed by switching from the mesoscopic H structure to a science chip Z structure, while keeping the base chip Z switched on.\;\scriptsize\circled{5}\small\; During evaporation to BEC the trap is decompressed once to avoid three-body collisions.}
	\label{fig:loading}
\end{figure}

This section will outline our loading scheme from generating a high flux of laser cooled atoms, to loading atoms into various layers of the atom chip, to high efficiency evaporative cooling. The total sequence can be divided into five steps as illustrated in Figure \ref{fig:loading}.

\subsection{Loading of the Chip MOT}
A 2D$^{+}$MOT\cite{Dieckmann1998} provides a beam of pre-cooled $^{87}$Rb atoms to the 3D-MOT. Its performance was characterized by measuring the fluorescence of probe light intersecting with the atomic beam in the 3D chamber\cite{Chaudhuri2006}. The magnitude of the fluorescence signal was used to find the optimal cooling light detuning of -18\,MHz and magnetic field gradient of 19.8\,G/cm. The magnetic field coils were adjusted individually to maximize flux through the differential pumping stage and compensate for small inequalities in laser power balance stemming from the mirror MOT configuration. The performance of the 2D$^{+}$MOT does not saturate at the total cooling laser power of 120\,mW that is currently available in the setup.

The longitudinal velocity profile of the atoms can be manipulated by changing the pusher-retarder power ratio (P/R) and the ratio between transverse and axial cooling power (T/A)\cite{Chaudhuri2006}. The capture velocity of the chip MOT was found to be limited to 30\,m/s by simulating the capture process through solving the equation of motion of the atoms numerically. This relatively low value is caused by the small beam diameters of 18\,mm and the fact that the magnetic field only has a true quadrupole shape in the vicinity of the trap center. Therefore, the velocity profile of the source was tailored to the capture performance of the chip MOT. The mean velocity and velocity spread of the atomic beam are 22\,m/s each, for power ratios of \mbox{P/R = 4} and \mbox{T/A = 18}.

The 3D-MOT features magnetic field gradients of $(B'_x, B'_y, B'_z) \approx (20, 20, 6)\,$G/cm and a cooling laser detuning of -20\,MHz. A total cooling laser power of 92\,mW is available, but the optimal MOT performance is achieved with 15\,mW in each of the four MOT beams. Optimized for highest atom number, the initial flux of atoms into the 3D-MOT is $1 \times 10^9$\,atoms/s and the MOT saturates at $2.5 \times 10^9$\,atoms after 4\,s. However, the loading can also be optimized for shorter times, featuring higher initial flux while saturating at a lower total atom number. A typical MOT loading phase prepares about $1 \times 10^{9}$ atoms in 500\,ms. This phase can be reduced to as little as 150\,ms without a significant decrease in BEC performance, as the transfer to the magnetic trap is the most critical step of the loading process. 

\subsection{Compressed MOT, Molasses Cooling and Optical Pumping}
After MOT loading, the center of the cloud is located at a distance of approximately 3\,mm from the chip's surface. Several measures are employed to optimize loading into the first magnetic trap which is centered only 500\,\textmu m away from the atom chip.

Firstly, the cloud is compressed by reducing the magnetic field gradient to 3\,G/cm and increasing the detuning to -48\,MHz. The atoms are then shifted towards the chip by adjusting the bias field. The entire process lasts approximately 40\,ms. Since the cloud is still around 1\,mm in size, it is not possible to match the cloud's and the magnetic trap's positions perfectly without moving a portion of the atoms into the chip.

Secondly, the atoms are subjected to a brief period of polarization gradient cooling in an optical molasses\cite{Chu1985}. To this end, the magnetic field is switched off and light detuned by -116\,MHz from the cooling transition is applied for 2\,ms. The final temperature of the cloud is 20\,\textmu K. Smaller molasses temperatures can not be obtained in our setup due to the inherently poor beam balance in the mirror MOT configuration.

Thirdly, the atoms are optically pumped into the ($F=2, \, m_\text{F}=2$) state by applying cooling light with a linear frequency sweep from -265\,MHz to -244\,MHz over 0.73\,ms. This state preparation pulse increases the number of atoms transferred to the initial magnetic trap by a factor of 3.

\subsection{Transfer to Initial Magnetic Trap}
Efficient transfer into the first magnetic trap depends on several parameters: co-location of atoms and trap, trap volume and mode matching. Mode matching is achieved by minimizing the change in entropy $\Delta S \geq 0$ and thereby heating caused by the transfer. Ideally, the phase space density (PSD) of the molasses cooled cloud,
\begin{equation}
n_0 \Lambda^3 = \exp \left(\frac{5}{2} + \gamma - \frac{S}{N} \right),
\end{equation}
should be conserved in the thermalized, magnetically trapped ensemble. Here, $n_0$ is the peak density, $\Lambda$ the thermal de\;Broglie wavelength, $\gamma=3/2$ the effective volume of a harmonic trap and $S/N$ the entropy per particle\cite{Pinkse1996}. We find that the harmonic approximation is valid for the cloud sizes and temperatures in question. So if we assume instantaneous transfer and that the initial cloud is normally distributed \cite{Ketterle1999} with width $\sigma$ and temperature $T$, we find that the optimal harmonic trap frequency is given by:
\begin{equation}
f = \frac{1}{2 \pi \sigma} \sqrt{\frac{k_\text{B} T}{m}},
\end{equation}
where $k_\text{B}$ is the Boltzmann constant and $m$ the particle mass\cite{Reinhard2009}. In our case, an optimal transfer would require trap frequencies around 6\,Hz which is not feasible in the presence of gravitational sag. Instead, we find optimal transfer into a IP type trap with trap frequencies (62, 58, 8.2)\,Hz, generated by the mesoscopic H structure, a base chip Z structure and the bias coil.

Overall, we observe an increase in temperature by a factor of 2 as well as a transfer efficiency of 25\% for saturated MOTs. The efficiency deteriorates with the number of atoms, i.e.\ smaller clouds can be transferred more efficiently. The highest number of atoms transferred is $2 \times 10^8$.

In summary, we are limited by the spatial mismatch of cloud and trap center, the inability to use the correct trap frequencies and the depth of the magnetic trap. Both the spatial mismatch and the gravitational sag can be circumvented by operating the experiment in microgravity, where shallow enough traps can be used at larger distances from the atom chip.

\subsection{Transfer to Final Magnetic Trap}
Once the atoms are confined in a magnetic trap and have reached thermal equilibrium, they are transferred adiabatically to the final trap configuration in two steps. Firstly, the atoms are loaded into a superposition trap of the base chip Z structure and a science chip Z structure, by simultaneously switching off the mesoscopic H while switching on the science chip over 25\,ms. Afterwards, the trap is compressed by increasing the bias current and the atoms are pulled closer to the chip over 100\,ms. The final trap features trap frequencies of (1783, 1779, 20.6)\,Hz and the temperature increases to 180 \textmu K. The initial PSD is $10^{-5}$ at an elastic collision rate of 500\,Hz.

\subsection{Evaporative Cooling to BEC}\label{sec:evap}
After transfer to the final magnetic trap, the atoms are cooled towards the critical temperature for Bose-Einstein condensation by selectively removing atoms with more than the average energy from the trap using radio frequency (RF) photons\cite{Pritchard1989,Davis1995,Anderson1995}. To this end, an RF source is connected to a dedicated U-structure on the base chip. Starting from 18.8\,dBm, the output power is continually attenuated to \textless 1\,dBm at the end of the evaporation sequence.

\begin{figure}[t]
	\centering
		\includegraphics[width=0.7\textwidth]{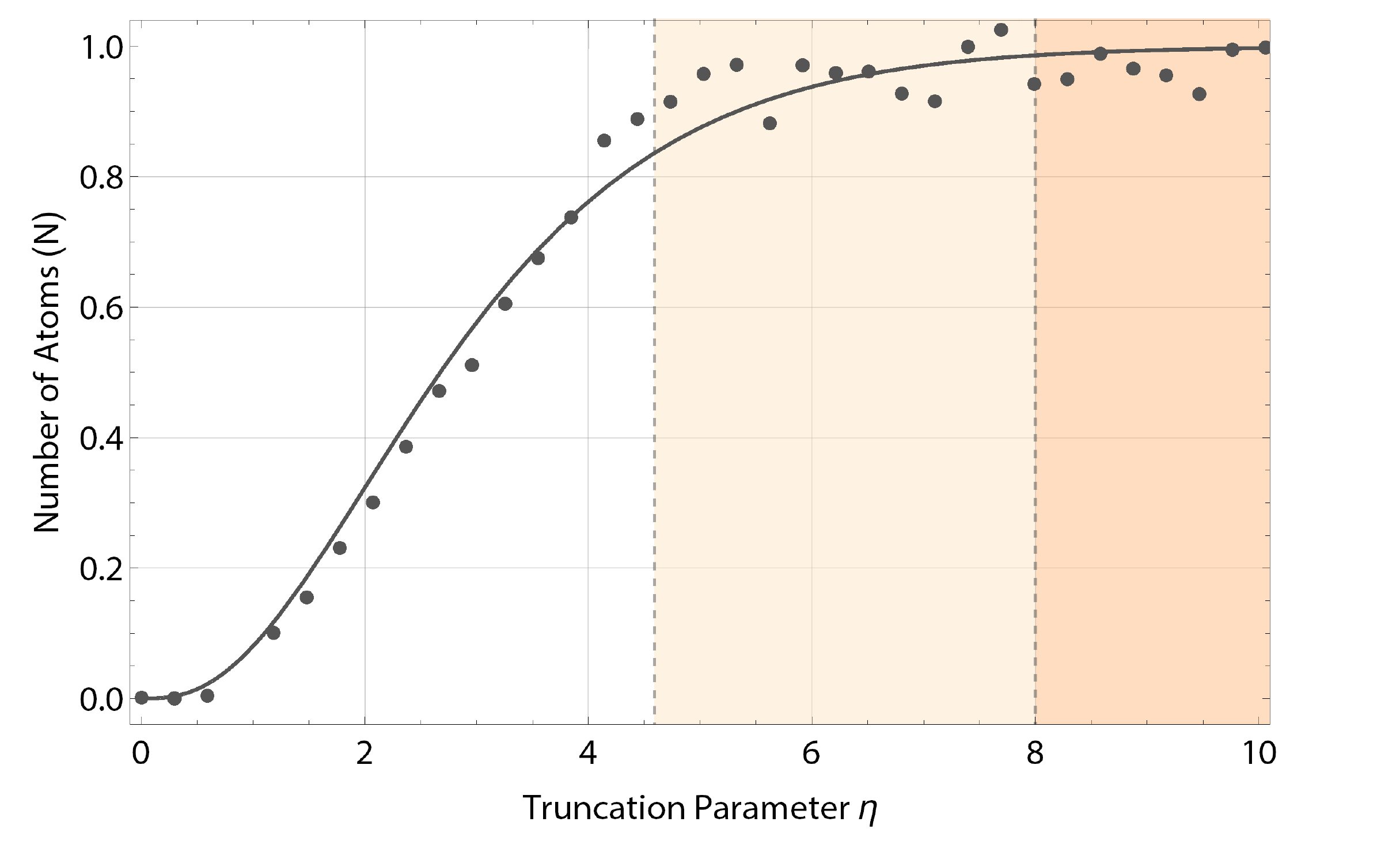}
	\caption{Cumulative atom number distribution over the truncation parameter $\eta = \epsilon/k_\text{B} T$. The solid line is a fit of Equation (\ref{eq:tdist}) to atom number data acquired by truncating the distribution at various energies. For efficient evaporative cooling, truncating at energies $>4.59\,\eta$ is desirable. For minimal atom loss the truncation energies should exceed $8\,\eta$. \newline The results are in good agreement with time-of-flight (TOF) measurements despite the fit not being perfectly representative of the data for energies larger than $4\,\eta$. The deviation stems from the fact that truncating at higher energies leads to cooling.}
	\label{fig:tdist}
\end{figure}

Since the initial temperature $T$ is much higher than $T_\text{c}$, the energy $\epsilon$ of the ensemble follows a Boltzmann distribution $f(\epsilon) = n_0 \Lambda^3\, e^{-\epsilon/k_\text{B} T}$, while the distribution of atoms in the trap is governed by the density of states $g(\epsilon) =  \epsilon^2 / 2 (\hbar \overline{\omega})^3$. Here $\overline{\omega}$ is the geometrical mean of the trapping frequencies. The total number of atoms can thus be obtained from
\begin{equation}
	N = \int_{0}^{\infty}{d\epsilon \; g(\epsilon) f(\epsilon)}
\end{equation}
and the number of atoms up to a given threshold energy $\epsilon_t$ can be expressed as\cite{Simonet2011}:
\begin{eqnarray}\label{eq:tdist}
	N_\text{t} &= N - \int_{\epsilon_t}^{\infty}{d\epsilon \; g(\epsilon) f(\epsilon)} \nonumber \\
	&= N \left[ 1 - e^{-\case{\epsilon_t}{k_\text{B} T}} \left(1 + \case{\epsilon_t}{k_\text{B} T} + \case{1}{2} \left(\case{\epsilon_t}{k_\text{B} T}\right)^2 \right) \right].
\end{eqnarray}
Applying an RF knife with frequency $f$ limits the total energy for trapped atoms to
\begin{equation}
	\epsilon_t = |m_\text{F}|\, h (f - f_0) = \eta\, k_\text{B} T
\end{equation}
by coupling atoms with higher energies to untrapped states and thus evaporating them from the trap. Here, $f_0$ is the resonance frequency at the bottom of the trap and $\eta$ is called the truncation parameter. The distribution (\ref{eq:tdist}) can be measured by truncating at various energies and recording the remaining number of atoms. Fitting Equation (\ref{eq:tdist}) to the data yields the total number of atoms, the trap bottom frequency and the temperature (see Figure \ref{fig:tdist}).

After truncation, the ensemble rethermalizes through elastic collisions and arrives at another energy distribution with lower temperature. To remove the same amount of energy over time, $\eta$ is held approximately constant and $f$ is ramped down exponentially. We break down the exponential frequency ramp into five linear ramps to be more flexible towards the conditions in each phase and optimize the efficiency of the cooling process step by step.

Generally, it is desirable to use a truncation parameter as high as possible to keep atom loss to a minimum. However, in the presence of other loss mechanisms the optimal truncation parameter and thereby maximal evaporation efficiency must be obtained from a comprehensive model of the process. The efficiency found experimentally for our optimized sequence is depicted in Figure \ref{fig:psd}a.

\subsubsection{Efficiency Model} 
\begin{figure}[htp]
	\centering
		\includegraphics[width=0.75\textwidth]{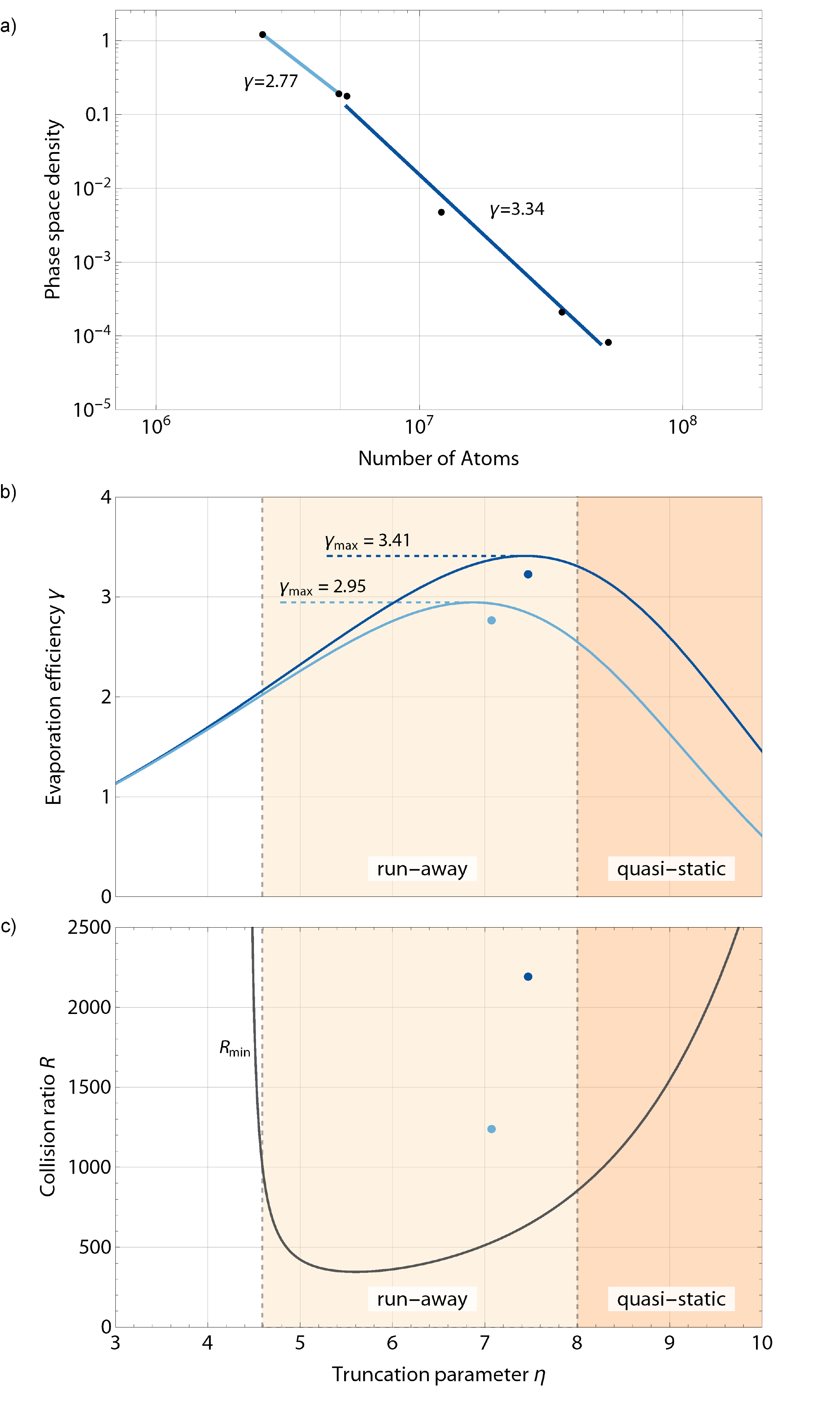}%
			\caption{Evaporation performance. Figure \ref{fig:psd}a shows the phase space density of the ensemble over the atom number after each of the linear RF ramps. The $\gamma$ factor has been averaged over the ramps in each trap configuration (dark blue and light blue). Figure \ref{fig:psd}b plots the modeled efficiency $\gamma(\eta, R)$ for the respective values of $R$ shown in Figure \ref{fig:psd}c. The attained efficiencies are close to the maximum values predicted by the model. The entire sequence is performed deep in the runaway regime bounded by $R_\text{min}$.}%
	\label{fig:psd}
\end{figure}

\setcounter{footnote}{5}
The figure of merit for efficient evaporative cooling is the ratio between the change in PSD to the change in atom number $N$ of the ensemble:
\begin{equation}
	\gamma = -\frac{\text{d}\, \ln(\text{PSD})}{\text{d}\, \ln(N)}.
\end{equation}
If we assume perfect forced evaporation, i.e.\ every atom with an energy higher than $\eta\, k_\text{B}T$ is immediately evaporated, there is a simple relation between the truncation parameter $\eta$ and $\gamma$
\begin{equation}\label{eq:gamma}
	\gamma(\eta, R) = \frac{3\, \alpha(\eta)}{1 - \lambda(\eta)/R } - 1,
\end{equation}
for a three-dimensional harmonic potential\cite{Ketterle1996}. Here, $\alpha$ is the ratio between a change in temperature and a change in atom number, $\lambda$ the ratio between the time constant of the evaporation and the elastic collision time and $R$ the ratio of elastic to inelastic collisions. The functions $\alpha(\eta)$ and $\lambda(\eta)$ are independent of $R$ and can be obtained from a suitable model\cite{Ketterle1996}. For arbitrary values of $\eta$ these functions need to be expressed in terms of incomplete gamma functions $P(n, \eta)$\footnote{These functions are related to the {Euler} gamma function {$\Gamma$} via ${P(a, \eta)} = \frac{1}{\Gamma(a)} \int_{0}^{\eta}{dt\, t^{a-1}\, e^{-t}}$ and ${R(n, \eta)}$ is given by ${R(a, \eta)} = \frac{P(a + 1, \eta)}{P(a, \eta)} = 1 - e^{-\eta} \frac{\eta^{a}}{\Gamma(a + 1)} \frac{1}{P(a, \eta)}.$} to reflect the truncated Boltzmann distribution of the energy\cite{Walraven1996,Luiten1996}:
\begin{eqnarray}\label{eq:trunc}
	\alpha(\eta)  &= \frac{1 + \eta - \case{P(5,\eta)}{\eta \, P(3,\eta) - P(4,\eta)} - 3 R(3,\eta)}{3 R(3,\eta) + 3 \left( 1 - \case{P(5,\eta)}{\eta \, P(3,\eta) - P(4,\eta)} \right) \Big(  1 - R(3,\eta) \Big)},	\nonumber \\
	\lambda(\eta) &= \sqrt{2} \Big[ 1 - 3 \Big( 1 - R(3,\eta) \Big) \, \alpha(\eta) \Big] \, \case{P(3,\eta)}{\eta \, P(3,\eta) - 4 P(4,\eta)} \, e^{\eta}.
\end{eqnarray}
Using equations (\ref{eq:gamma}) and (\ref{eq:trunc}), the evaporation efficiency $\gamma(\eta, R)$ for perfect forced evaporation in a three-dimensional harmonic potential has been plotted in Figure \ref{fig:psd}b for the two trap configurations in use and their respective value of $R$.

\subsubsection{Collision Rates}
Having derived analytical expressions for $\alpha(\eta)$ and $\lambda(\eta)$, it is evident from Equation (\ref{eq:gamma}), that in a system where $\eta$ can be chosen arbitrarily, the maximum efficiency $\gamma_{\text{max}}$ only depends on $R$, the ratio of good to bad collisions.

The good collisions are those between two trapped atoms in the same spin state which lead to rethermalization without atoms being lost from the trap. This process has the per atom rate $\Gamma_{\text{el}}$. The relevant inelastic processes are collisions with atoms from the background gas with rate $\Gamma_{\text{bg}}$ and inelastic three-body collisions with per atom rate $\Gamma_{\text{3-body}}$. This leaves us with the final ratio
\begin{equation}
R = \frac{\Gamma_{\text{el}}}{\Gamma_{\text{loss}}} = \frac{\Gamma_{\text{el}}}{\Gamma_{\text{bg}} + \Gamma_{\text{3-body}}}.
\end{equation}
Since $\Gamma_{\text{bg}}$ is generally constant over the evaporation process, maximizing $R$ means increasing the elastic collision rate as much as possible while staying dilute enough to not give rise to three-body collisions. For this reason, the trap is initially compressed to trap frequencies of (1783, 1779, 20.6)\,Hz to maximize the elastic collision rate. As the ensemble grows more and more dense, the trap is decompressed to trap frequencies of (701, 696, 23.3)\,Hz after four ramps of evaporation, at the onset of three-body collisions. These two trap configurations are color coded in dark blue and light blue, respectively, in Figure \ref{fig:psd}.

However, in the presence of other loss mechanisms, e.g.\ due to the limited trap depth, $R$ can more generally denote the ratio of the lifetime of the ensemble over the elastic collision time. For the data presented, the lifetime in the trap averaged over the entire sequence was measured to be 2.84\,s.

If the elastic collision rate is to stay constant or increase during evaporation (runaway evaporation), a lower limit for $R$ can be specified\cite{Ketterle1996}:
\begin{equation}
R_{\text{min}} = \frac{\lambda(\eta)}{\alpha(\eta) - 1}.
\label{eq:rmin}
\end{equation}
Figure \ref{fig:psd}c plots the theoretical bounds set by $R_{\text{min}}$ using equations (\ref{eq:trunc}) and (\ref{eq:rmin}), together with the measured values of $R$ for each step of the evaporation sequence. The runaway regime for a three-dimensional harmonic potential starts at $\eta > 4.59$. Beyond a truncation energy of $\eta = 8$, $R_{\text{min}}$ starts to increase rapidly as very few atoms populate the high energy tail of the distribution. Operating in this quasi-static regime offers minimal atom loss but generally requires lifetimes \textgreater 30\,s at the trap frequencies we employ.

While we are operating close to the maximum efficiency as predicted by the model, we are clearly limited by the lifetime of the ensemble. Higher efficiencies and atom numbers may be obtained if the vacuum quality can be improved and the evaporation sequence can be maintained in the quasi static regime.
\section{Discussion and Comparison}\label{sec:comparison}
\begin{figure}[htp]
\centering

\includegraphics[width=0.78\textwidth]{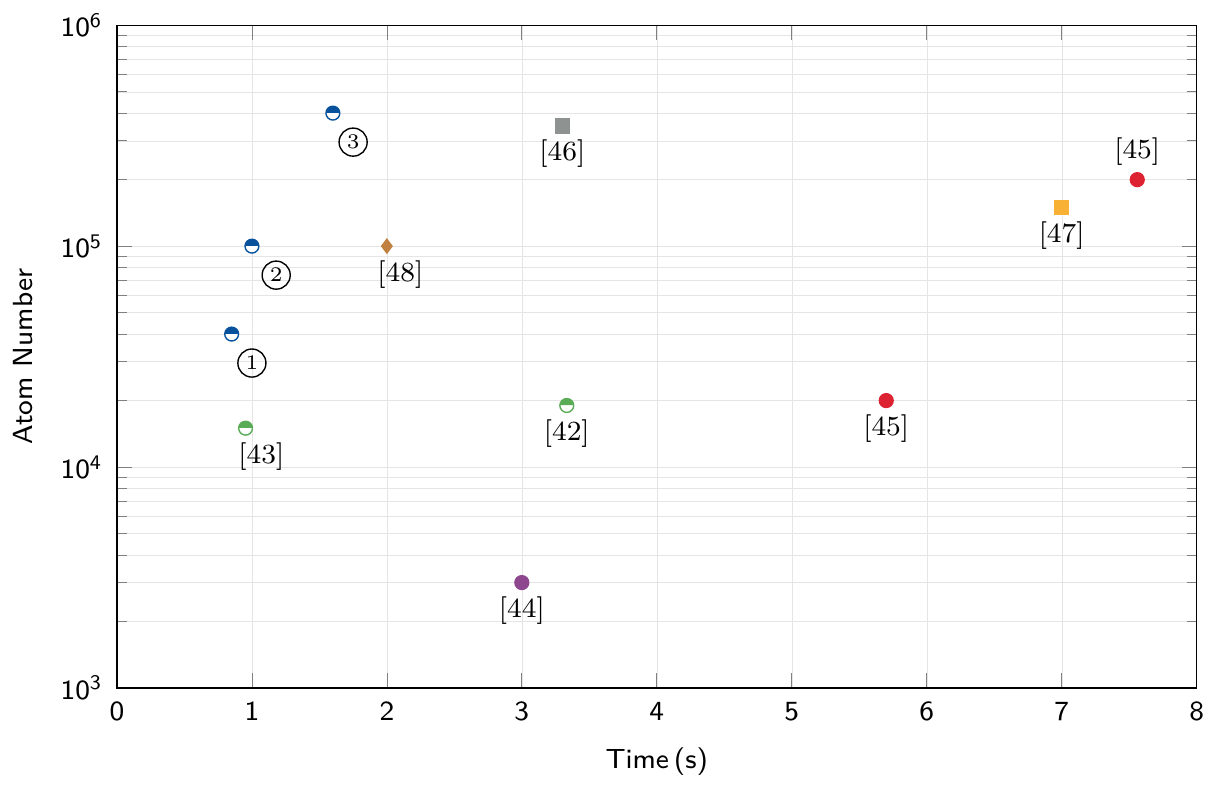}
\caption{Comparison of the fastest BEC machines. Circles denote atom chip based experiments\cite{Farkas2010,Farkas2014,Horikoshi2006,Extavour2006}, squares indicate experiments using dipole traps\cite{Kinoshita2005,Clement2009}. The diamond symbol indicates a Sr experiment reported in\cite{Stellmer2013}. Semi-filled symbols mark compact and transportable setups. The results of this work are represented by three cases, \scriptsize\circled{1} - \circled{3}\normalsize.}
	\label{fig:compare}
\end{figure}

Since the first demonstration of BEC in 1995, the cycle time of generating quantum degenerate gases has decreased steadily. The fastest BEC machines published to date are compared in Figure \ref{fig:compare}. These experiments either employ an atom chip (circles) or a dipole trap (squares) for fast and efficient evaporation. Recently, BEC machines are not only getting faster but have also become much more compact\cite{Zoest2010,Farkas2010}. They have since crossed the divide from lab experiments to mobile and transportable devices (semi-filled symbols).

The fastest previously reported apparatus features a repetition rate of 1\,Hz with $1.5 \times 10^4$ atoms in the condensed phase, using an atom chip\cite{Farkas2014}. The fastest BEC in a dipole trap was produced in 3.3\,s with an atom number of $3.5 \times 10^5$\cite{Kinoshita2005}. To study the overall performance of our setup, we optimized the BEC production for three scenarios:  fast BEC production \scriptsize\circled{1}\normalsize, BEC at 1\,Hz repetition rate \scriptsize\circled{2} \normalsize and highest atom number in the BEC \scriptsize\circled{3}\normalsize. The shortest production time can be achieved by reducing the MOT loading time to 350\,ms and the duration of the evaporation to 450\,ms. Within a total time of 850\,ms the apparatus is able to produce BECs of 4 $\times$ 10$^4$ atoms. For a direct comparison with the previously fastest BEC machine, we optimized the atom number at a production rate of 1\,Hz. With a MOT loading time of 450\,ms and a duration of the evaporation of 500\,ms, we are able to produce ensembles of $1 \times 10^5$ atoms. The experimental sequence for the highest atom number in the BEC was presented in the previous chapter.

These results can be improved further, both in terms of speed and atom number. The biggest handicap for the data presented was the vacuum quality and thereby the lifetime of the atoms. Improving the vacuum quality could reduce the MOT loading time to as little as 100\,ms. A similar improvement can also be achieved by increasing the 2D cooling laser power. An increase in lifetime would also lead to higher evaporation efficiencies and thus higher atom numbers. Another approach to increase the atom number in the BEC is to improve the mode matching of the laser cooled atoms with the initial magnetic trap, resulting in a higher initial PSD. This could be achieved by reducing the molasses temperature further or by operating the experiment in microgravity, where the optimal mode matching conditions can be met thanks to the absence of gravitational sag. Consequently, producing BECs of $10^6$ atoms at a 1\,Hz rate seems feasible if these technical issues are addressed.

While our setup produces the highest atom number overall among the fastest BEC machines, its flux of condensed atoms is also on par with the best lab-based devices\cite{Streed2006,Lin2009}. As a result, the advances presented in this paper are not only of interest for mobile applications but any experiment that benefits from high repetition rates. Therefore, we anticipate a vast range of applications for our source: The reduction of the BEC production time by one order of magnitude immediately leads to a significant improvement on the sensitivity of atom interferometers, a field where great efforts are made to design machines with high data rate\cite{Rakholia2014}. It is also of great interest for studying non-classical correlations in BECs and hence for quantum optics experiments with atoms in general. State-of-the-art quantum enhanced magnetometers\cite{Vengalattore2007,Muessel2014,Ockeloen2013} would be three times to one order of magnitude more sensitive, reaching the sub-pT/$\sqrt{\text{Hz}}$ regime. Any experiment requiring large statistics, for example to monitor correlations in quantum many-body systems \cite{Schumm2005,Langen2013}, would dramatically benefit from such a source by dividing the total time to take the data by a factor 3 to 10. However, the use of the source is not restricted to fundamental physics, but provides a new tool for earth observation with improved accuracy in geodesy and geophysics.

The ability to demonstrate such brilliant BEC sources for mobile devices is at the heart of current precision sensor proposals. Indeed, these experiments rely on a dramatic increase of the interferometry times compared to state-of-the-art realizations. Recent proposals for testing fundamental laws or detecting gravitational waves in atomic fountains\cite{Dimopoulos2007}, drop towers\cite{Zoest2010}, parabolic flights\cite{Geiger2011} or space missions\cite{Hogan2011, Altschul2015, Aguilera2014} rely on total times of flight of several seconds. A thermal cloud of typically $10^{9}$ atoms at a temperature of 1\,\textmu K  reaches sizes of a few tens of mm after such durations. In contrast, a BEC of $10^6$ atoms would expand to only a few mm as confirmed by pioneering experiments\cite{Dickerson2013, Muentinga2013}. This feature makes degenerate atomic ensembles an exquisite source for high-contrast interferometry experiments on macroscopic timescales. At these scales, the mean field effects associated with substantial dephasing in trapped or short-time atom interferometers are dramatically reduced\cite{Hartwig2015}. In fact, expanding a BEC from a typical size of 50\,μm to 1\,mm suppresses the effects of interactions by 4 to 5 orders of magnitude.

Moreover, using our experimental design in precision interferometry offers some intrinsic advantages related to BEC operation. Indeed, the small momentum width of degenerate atoms is a clear gain with respect to the efficiency of atom-light-interaction based beam splitters, since the phase sensitivity of an atom interferometer is fundamentally limited by the fidelity of momentum transfer in an atom-based sensor\cite{Szigeti2012}. This holds true even more for magnetic and optical lensing techniques, where the mode quality of BECs is mandatory to reach extremely low expansion rates necessary for the reduction of the most relevant systematics. Operating with thermal ensembles results in orders of magnitude larger clouds with dramatically increased aberration effects\cite{Kovachy2014}.

The advantages stated above would be compromised if the cycle times would be identical to the ones of traditional BEC experiments. With the current device, it is possible to integrate down to a desired target accuracy quite efficiently. For example, the state-of-the-art accuracy ($10^{-13}$) for WEP tests can be obtained after 4 hours of integration for interferometry times of $T = 2$\,s.

\section{Conclusion}\label{sec:conclusion}
We studied the performance of a new generation of compact BEC machines relying on the use of atom chips. Despite the significant miniaturization, the device is able to deliver a flux of quantum degenerate $^{87}$Rb atoms comparable to the best laboratory experiments. Depending on the experimental need, several scenarios with different atom numbers and cycle times could be performed. A detailed study of the route to BEC, from background atomic vapor to the final chip trap, was presented with key device-specific techniques illustrated. The unprecedented compactness of the experimental payload confirms the possibility to embark quantum gas sensors in microgravity platforms such as drop towers, parabolic flights or sounding rockets. Moreover, it announces the era of utilizing atom interferometers in space in extended free fall. Several recent proposals anticipate an outstanding sensitivity of space-borne quantum sensors testing fundamental theories and principles.

\ack
The authors would like to thank Jakob Reichel and Friedemann Reinhard for valuable discussions and support with the design and construction of the atom chip. This project is supported by the German Space Agency (DLR) with funds provided by the Federal Ministry for Economic Affairs and Energy (BMWi) due to an enactment of the German Bundestag under grant numbers DLR 50 1131-1137 (QUANTUS-III). This work was also supported in part by the Centre for Quantum Engineering and Space-Time Research (QUEST). J.R.\ acknowledges support by the Hannover School for Laser, Optics and Space-Time Research (HALOSTAR). The authors also wish to thank the German research foundation DFG for funding the collaborative network SFB geo-Q.

\section*{References}
\bibliographystyle{iopart-num}
\bibliography{fastbecbib}

\end{document}